# A Dynamical Investigation of the Proposed BD +20 2457 System

J. Horner[1,2], R. A. Wittenmyer[1,2], T. C. Hinse[3], & J. P. Marshall[4]

[1] School of Physics, University of New South Wales, Sydney, NSW 2052, Australia
[2] Australian Centre for Astrobiology, University of New South Wales, Sydney, NSW 2052, Australia
[3] Korea Astronomy and Space Science Institute, 776 Daedeokdae-ro Yuseong-gu 305-348 Daejeon, Korea
[4] Departamento de Física Teórica, Facultad de Ciencias, Universidad Autónoma de Madrid, Cantoblanco, 28049 Madrid, Spain

**Abstract**
We present a detailed dynamical analysis of the orbital stability of the BD +20 2457 system, which features planets or brown dwarfs moving on relatively eccentric orbits. We find that the system exhibits strong dynamical instability on astronomically short timescales across a wide range of plausible orbital eccentricities, semi-major axes, and inclinations. If the system truly hosts massive planets or brown dwarfs, our results suggest that they must move on orbits significantly different to those proposed in the discovery work. If that is indeed the case, then it is likely that the best-fit orbital solutions for the proposed companions will change markedly as future observations are made. Such observations may result in the solution shifting to a more dynamically-stable regime, potentially one where stability is ensured by mutually resonant motion.

## 1. Introduction

In recent years, the number of planets announced orbiting around other stars has increased dramatically. Where once single discoveries were the norm (e.g. Mayor & Queloz, 1995; Marcy & Butler, 1996; Butler & Marcy, 1996), systems featuring multiple planets are now being discovered ever more frequently (e.g. Butler et al., 1999; Lovis, 2011; Robertson et al., 2012a, b; Wittenmyer et al., 2012a), often around stars far different from our own Sun (e.g. Wolszczan & Frail, 1992; Beuermann et al., 2010; Muirhead et al., 2012; Sato et al., 2012, 2013).

One common feature of these newly discovered exoplanetary systems is that the vast majority are found by indirect means – such as the Radial Velocity and Eclipse Timing Variations techniques (e.g. Perryman, 2011). In essence, these methods look for periodic variations in an observable property of a star (its line-of-sight motion, in the case of radial velocity observations, and the timing of eclipses between a close binary star system for the eclipse timing technique), and attempt to explain any variations detected as being the result of the influence of massive unseen companions.

For most systems, the planets considered are well separated or of low enough masses that interactions between them can be ignored, and the data can be fit with Keplerian (non-interacting) orbits. Whilst this technique is perfectly reasonable when only a single planet is thought to orbit a given star, it can fail when attempting to fit multiple massive planets, leading to proposed orbital solutions featuring planets that strongly interact, or even collide, with one another on very short timescales (e.g. Horner et al., 2011, 2012a; Wittenmyer et al., 2012b, 2013a).

In this light, it is clearly important to complement the orbital fitting of observational data with dynamical simulations that check whether the orbital fits obtained are reasonable (e.g. Goździewski, K., Konacki, M. & Migaszewski, C, 2006; Goździewski, Migaszewski & Musieliński, 2008; Veras & Ford 2010; Wittenmyer et al., 2013b). Such simulations have been used to better constrain the orbits of a number of recently discovered exoplanetary systems (e.g. Robertson et al., 2012a; Wittenmyer et al., 2012a), and can show how solutions that would otherwise by highly unstable can be stabilised by the influence of mutual mean motion resonance between the candidate planets (e.g. Horner et al., 2012b; Robertson et al., 2012b). At the same time, such simulations can also reveal systems for which the proposed planets simply are not dynamically feasible (e.g. Horner et al., 2011, 2012a), suggesting that further observations are necessary before conclusions can be drawn on the presence (or absence) of planets in a given system.

The BD +20 2457 system (Niedzielski et al., 2009) features two massive companions (most likely brown dwarfs) orbiting an evolved massive primary (a K-giant star almost three times the mass of the Sun). The candidate companions were announced as part of the Penn State–Toruń Planet Search, on the basis of 37 individual radial velocity observations obtained with the 9.2 m Hobby–Eberly Telescope over a period of 1833 days. The orbits proposed for the companions are moderately eccentric, and are sufficiently tightly packed that they may allow the proposed objects to experience strong mutual perturbations. As such, we have performed a detailed dynamical study of the system, to examine whether the candidate companions are dynamically feasible on their proposed orbits. In section two, we briefly describe the methodology with which we examine the dynamical stability of the proposed BD +20 2457 system, before presenting the results of that study in section three. Finally, we discuss our results and present our conclusions in section four.

## 2. Dynamical Simulations of Exoplanetary Systems

In order to study the dynamical feasibility of these recently proposed multiple planet systems, we followed a now well-established route (e.g. Marshall et al., 2010; Horner et al., 2011, 2012a, b; Wittenmyer et al., 2012a, b). We use the *Hybrid* integrator within the *n*-body dynamics package MERCURY (Chambers, 1999) to perform a series of integrations following the dynamical evolution of the chosen planetary systems for a period of 100 Myr, or until one or other of the planets therein is removed from the system as a result of collision (between the planets, or the planet and the central star) or ejection.

To study the dynamical evolution of the BD +20 2457 system, we carried out a main suite of 126,075 integrations, considering the scenario where the planets move on co-planar orbits. In those integrations, we held the initial orbit of the innermost planet fixed, with its nominal best-fit orbital elements. We then placed the outer of the two planets on initial orbits that ranged across the full ±3 sigma uncertainties in that planet's best-fit semi-major axis (*a*), eccentricity (*e*), longitude of periastron (omega) and mean anomaly (*M*). We considered 41 different initial values of semi-major axis and eccentricity for the outermost planet, each distributed evenly across the ±3 sigma uncertainties in those elements. At each of these *a-e* locations, we considered 15 different values of the longitude of periastron and 5 different values of mean anomaly, giving us a four-dimensional grid of 41x41x15x5 simulations (*a-e-omega-M*). This allowed us to plot the mean lifetime of the planetary system as a function of the semi-major axis and eccentricity of the outermost planet, as described previously in (e.g. Horner et al., 2011; 2012a, b).

In addition to these main runs, we also considered the influence of the mutual orbital inclination of the two planets. Subsidiary runs, at a lower resolution (21x21x5x5 in *a-e-omega-M*, for a total of 11025 runs per system) were carried out with the orbit of the outermost planet initially inclined to that of the innermost by 5, 15, 45, 135 and 180 degrees – again following our earlier work (e.g. Horner et al., 2011; Wittenmyer et al., 2013a, b).

## 3. BD +20 2457

The proposed companions of the K giant star BD +20 2457 (Niedzielski et al., 2009) are most likely brown dwarfs, rather than planets – with minimum masses of 21.4 and 12.5 times that of Jupiter. We have adopted a host-star mass of 2.8 solar masses as given in Niedzielski et al. (2009); we note that Mortier et al. (2013) report a significantly different value of 1.06±0.21 solar masses. Their semi-major axes (1.45 and 2.01 AU) and eccentricities (0.15 and 0.18) are such that their nominal best-fit orbits (detailed in Table 1) approach one another closely (the innermost object has an apastron distance of 1.67 AU, and the outermost a periastron distance of 1.65 AU) – a result that suggests the system might be extremely unstable[1]. As can be seen in Figure 1 (which shows the best-fit orbits proposed in Niedzielski et al., 2009), the orbits are currently oriented such that they merely approach one another relatively closely, rather than actually intersecting. However, unless those orbits are mutually resonant, their arguments of periastron would be expected to be precess at different rates, leading to mutually encountering orbits on relatively short timescales. However, the orbit of the outermost body as proposed in the discovery work features large uncertainties, which could clearly allow dynamically stable solutions to be found somewhere within the ±3 sigma uncertainties on those values.

---

[1] Indeed, mutually crossing orbits, and those that approach one another closely, are almost always dynamically unstable – unless the objects involved are protected from close encounters by the influence of mean-motion resonances (as is seen in our own Solar system for the Jovian and Neptunian Trojans, e.g. Horner & Lykawka, 2010; Horner et al., 2012c).

|  | BD +20 2457 b | BD +20 2457 c |
|---|---|---|
| Semi-Major Axis (au) | 1.45 | 2.01 ± 0.36 |
| Eccentricity | 0.15 ± 0.03 | 0.18 ± 0.06 |
| Mass[2] ($M_J$) | 21.42 | 12.47 |
| Omega (°) | 207.64 ± 21.99 | 126.02 ± 16.54 |
| $T_0$ (MJD) | 54677.03 ± 28.19 | 53866.9 ± 27.99 |

**Table 1:** The orbits of the two candidate companions to BD +20 2457, as detailed in Niedzielski et al., 2009. In that work, no uncertainties were provided for the semi-major axes of the two candidates, and so in this work, we use an uncertainty taken from http://exoplanets.org on 31st July 2012 for the semi-major axis of BD +20 2457 c.

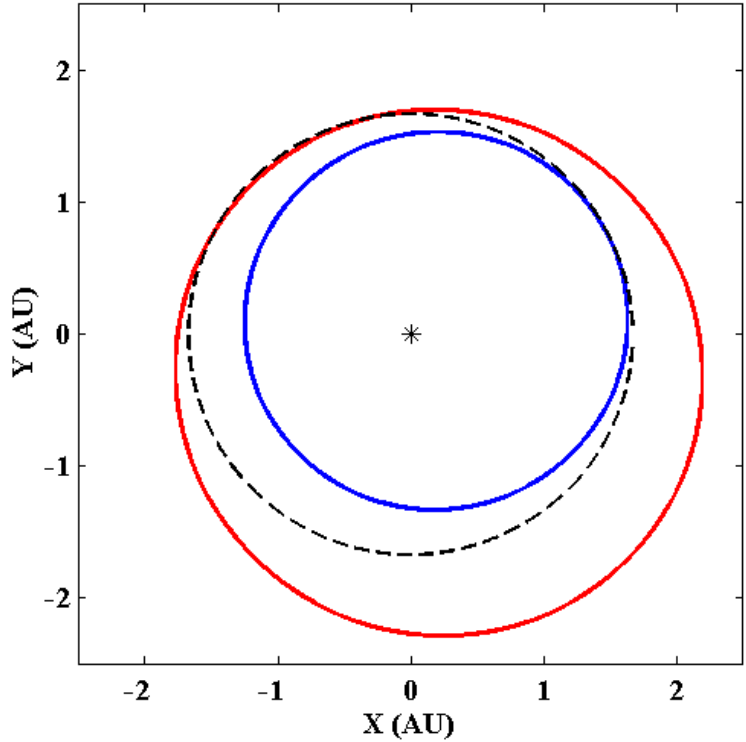

***Figure 1**: The best-fit orbital solutions for BD +20 2457 b (blue) and BD +20 2457 c (red), as proposed in Niedzielski et al., 2009. The location of BD +20 2457 is marked as a star. The dashed black circle, of radius 1.67 AU, shows the apastron distance of BD +20 2457 b, and serves to highlight the fact that, at periastron, BD +20 2457 c approaches within that distance. Unless the orbits of the two planets are mutually resonant, such a solution will inevitably eventually result in strong mutual encounters between the two objects, destabilising the system.*

---

[2] The mass quoted here is the *minimum* mass for the planets (*m* sin *i*) – the mass derived assuming that the companions orbit in the same plane as our line of sight. If the companion orbits are inclined to our line of sight by an angle *i*, then the true mass of the companions will be larger than this minimum value.

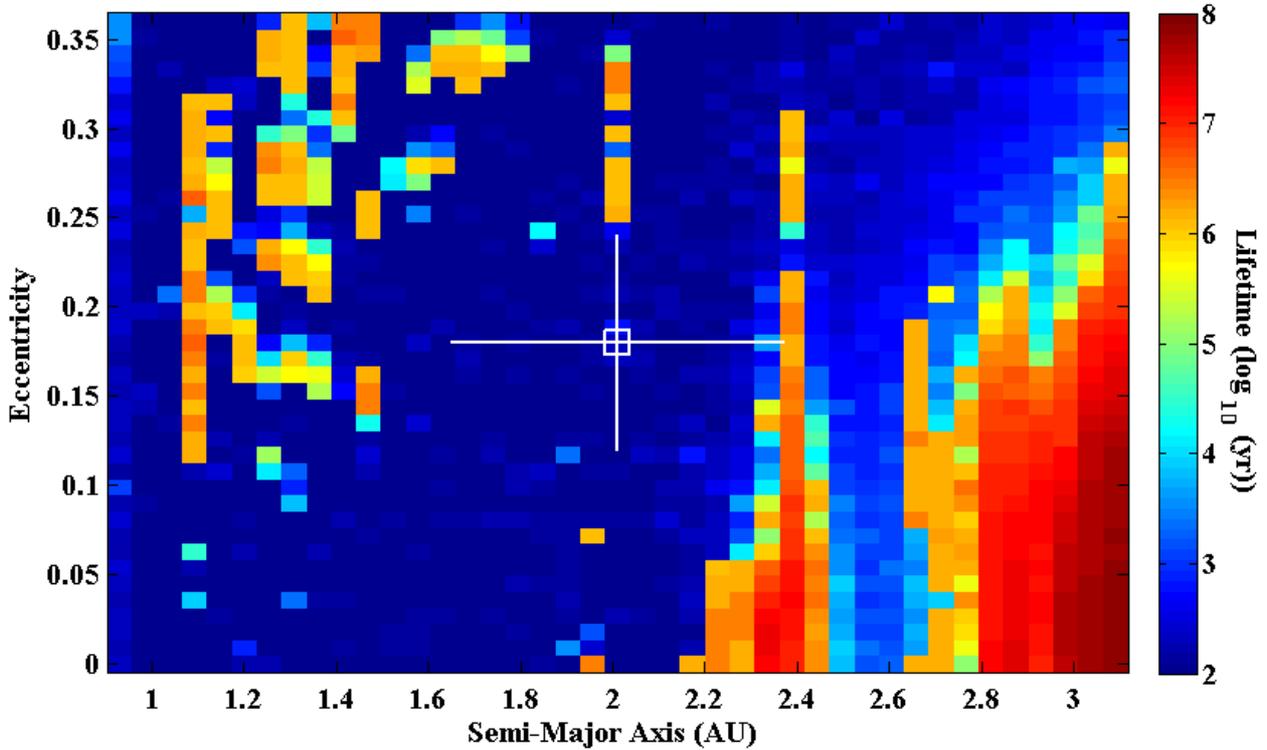

***Figure 2***: *The dynamical stability of the orbit of BD +20 2457 c, as a function of its orbital semi-major axis and eccentricity. The nominal best-fit orbit is located within the hollow square, with the ±1 sigma errors on that value being denoted by the white lines that radiate from that point. As a result of the large uncertainties in the orbit of BD +20 2457 c, the plot covers orbits that are both wholly interior to that of BD +20 2457 b, and wholly exterior to that planet, together with a wide variety of solutions where the orbits of the proposed planets would cross one another. For reference, the nominal best-fit orbit proposed for BD +20 2457 b is located at a = 1.45 AU, e = 0.15. The entire region within ±1 sigma of the best-fit orbit for BD +20 2457 c is highly dynamically unstable, with typical mean lifetimes of between 100 and 1,000 years.*

Figure 2 shows the results of our dynamical simulations of the BD +20 2457 system, for a scenario featuring co-planar orbits for BD +20 2457 b and c. It is immediately apparent that the great majority of the allowed solutions for the system are extremely dynamically unstable – on timescales of just a few hundred years. This broad instability includes all solutions within ±1 sigma of the nominal best-fit orbit in both eccentricity and semi-major axis. Interestingly, however, a narrow strip of stability can be seen at the nominal best-fit semi-major axis, for eccentricities greater than ~0.25. This is the result of the 5:3 mean-motion resonance, which is located at 2.038 AU when the orbit of BD +20 2457 b is located at its nominal best-fit value of 1.45 AU. The stabilising influence of the 2:1 mean-motion resonance can be seen around 2.3 AU, with orbits just exterior to the location of that resonance offering stability up to eccentricities equal to (and in excess of) the nominal best-fit value – a result entirely compatible with our earlier work. The 5:2 mean-motion resonance (at 2.67 AU) marks the inner edge of a broader region of dynamical stability, which is most strongly pronounced to the right of the location of the 3:1 mean-motion resonance (at 3.02 AU). Interestingly, a smattering of dynamically stable solutions can be seen at, and just interior to, the location of the nominal best-fit solution for the semi-major axis of BD +20 2457 b (at 1.45). These stable, yet mutually crossing, solutions are once again the result of the protective influence of a number of mutual mean-motion resonances between the two objects (e.g. the 1:1 MMR, at 1.45 AU, and the 2:3 MMR, at 1.1 AU, along with other, higher order resonances).

Given the extremely wide range of parameter space for which we have tested the dynamics, it is reasonable to ask how well those configurations match the observational data. As noted by Anglada-Escude et al. (2013) and Marsh et al. (2014), the parameter distributions are highly correlated, and the range shown in Figure 2 may include regions far more than 3σ from the best fit. We therefore computed the $\chi^2$ for each of the 126,075 systems tested (with 60 m/s jitter added in quadrature after Niedzielski et al., 2009). The results are shown in

Figure 3, on the same scale as Figure 2 for ease of comparison. We find that the islands of stability from Figure 2 lie in regions highly disfavoured by the data – with reduced $\chi^2$ greater than 5.

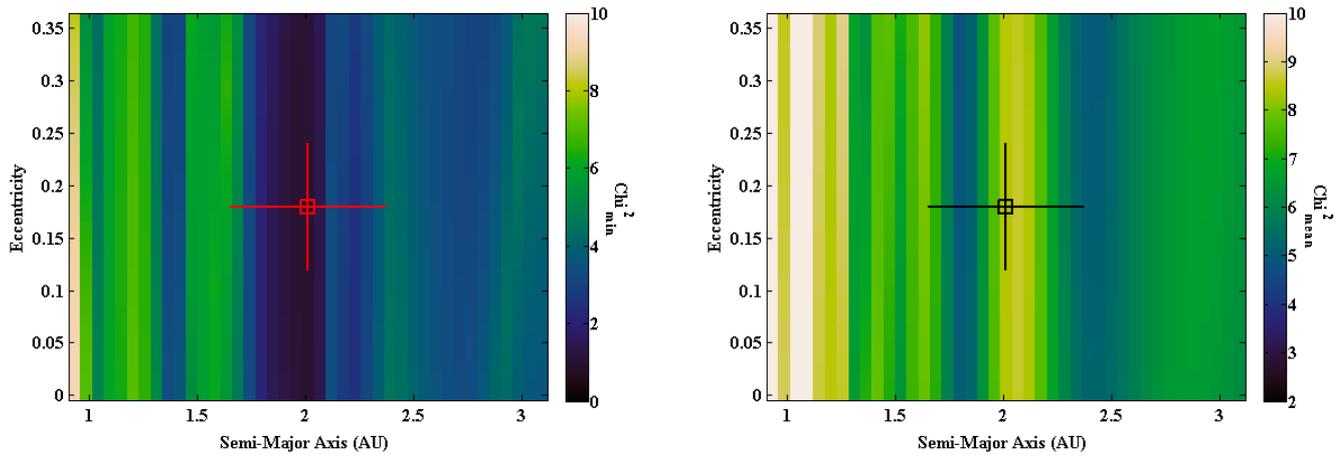

**Figure 3.** *Chi-square distribution of the 126,075 system configurations tested. The parameter space is divided into squares exactly as in Figure 2, where each square represents 75 individual "clones." Left panel: The minimum reduced $\chi^2$ of the 75 individual solutions is shown for each small square. Right panel: The mean reduced $\chi^2$ of the 75 individual solutions is shown for each small square. It is noteworthy that the islands of stability that are seen in figure 2 all lie in regions that are strongly disfavoured by the observational data.*

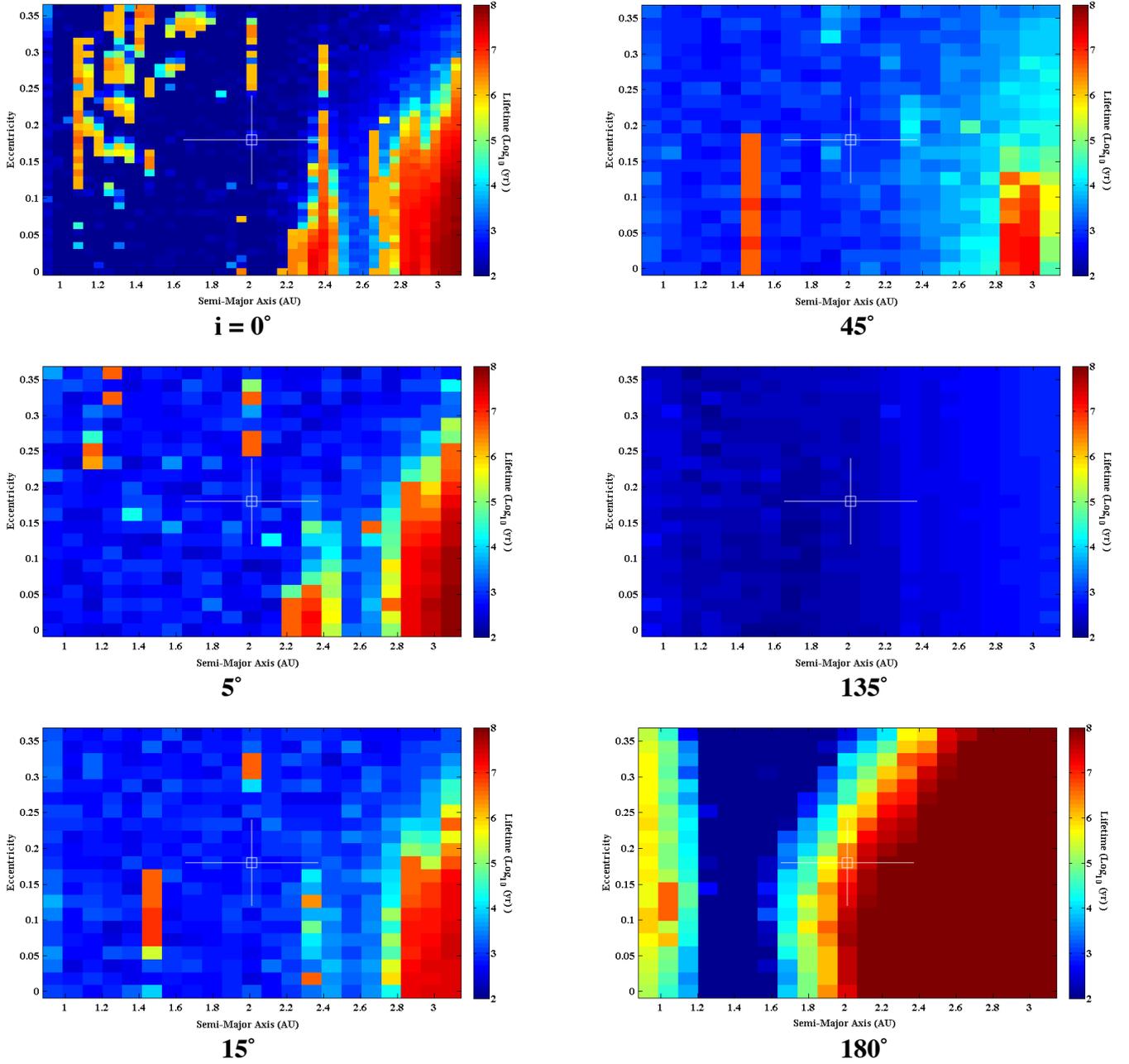

*Figure 4:* *The dynamical stability of the proposed BD +20 2457 planetary system, as a function of the mutual inclination between the orbits of BD +20 2457 b and BD +20 2457. The plots show the stability for mutual inclinations of 0 degrees (top left; see also Figure 2), 5 degrees (centre left), 15 degrees (lower left), 45 degrees (upper right), 135 degrees (centre right) and 180 degrees (lower right). The colour scale is the same across all panels, ranging from mean lifetimes of $10^2$ years (dark blue) to $10^8$ years (dark red).*

Figure 4 shows how the dynamical stability of the proposed BD +20 2457 system varies as a function of the mutual inclination between the orbits of the two proposed companion bodies. The moderately inclined scenarios (5 and 15 degrees, middle-left and lower-left hand panels) exhibit much the same features as the co-planar case discussed above – a broad region of instability around the nominal best-fit orbit, small regions of stability resulting from the influence of mutual mean-motion resonances, and a broader stable region towards larger semi-major axes. By the time the two objects have a mutual inclination of 45 degrees, only two regions of stability remain – the first, at low-to-moderate eccentricities, at 1.45 AU (the 1:1 MMR), and the second, again at low-to-moderate eccentricities, between 2.8 and 3 AU (i.e. between the 8:3 and 3:1 MMRs at 2.79 and 3.02 AU, respectively). For mutually retrograde orbits (the lower right hand panel), a wide variety of stable solutions are allowed, although mutually crossing solutions remain highly dynamically unstable. This result is not unexpected - such retrograde solutions are almost always highly stable unless they feature mutually crossing orbits (e.g. Eberle & Cuntz, 2010; Horner et al., 2011, 2012b, Wittenmyer et al., 2013a, b).

## 3.1 Verifying the orbital solution

We have found that mutually resonant solutions exist for the candidate planetary system orbiting BD +20 2457 that allow stability on timescales of millions of years – albeit at orbital eccentricities and semi-major axes relatively well removed from the best-fit solutions proposed in the discovery work. Since the host star is a highly evolved K giant (log g = 1.77± 0.19; Mortier et al 2013), the radial-velocity jitter is quite large; Niedzielski et al. (2009) included 60 m/s of jitter in their fitting. Additionally, when data are sparse or have large uncertainties, the determination of Keplerian orbital parameters is extremely difficult and can lead to degeneracies (Anglada-Escudé et al. 2010, Wittenmyer et al. 2013b).

As an additional check on the orbital parameters, we therefore re-fit the Niedzielski et al. (2009) radial-velocity data. We used a genetic algorithm to sample an extremely wide parameter space in search of a truly global best fit. This technique has frequently been used for systems with highly uncertain parameters (e.g. Tinney et al. 2011, Wittenmyer et al. 2012b, Horner et al. 2012b). The genetic algorithm generates a random population, whose members are described by the set of parameters to be solved for. The user defines an allowed range for each parameter, and the "genotype" of each population member is chosen randomly from within that range.

Once this process is complete, the $\chi^2$ merit function is computed for each member (set of planetary parameters), and that $\chi^2$ corresponds to its "fitness" in the population: models resulting in lower $\chi^2$ are more fit. As in biological evolution, recombination and mutations occur, and the fittest population members have a higher probability of reproducing for the next generation. In this manner, the genetic algorithm slowly converges to a global $\chi^2$ minimum by sampling all allowed parameter space. For the BD+20 2457 system, we used a population of 1000 models, allowed to evolve until the change in total $\chi^2$ was less than $10^{-3}$ between successive generations. A total of 50000 such iterations were performed, each one resulting in a set of parameters and a $\chi^2$ for a 2-planet model. The inner planet was allowed to take an orbital period in the range 300-500 days and an eccentricity between 0.0 and 0.4, whilst the outer planet was allowed a period of 500-800 days and an eccentricity of 0.0 to 0.4. From more than $5 \times 10^7$ individual trials, we found that the global best-fit solution is indeed in agreement with Niedzielski et al. (2009).

## 4. Conclusions

Our results suggest that the massive companions proposed to orbit the evolved giant BD +20 2457 do not exist on the nominal best-fit orbits suggested in that work. Orbits within ±1 sigma of the best solutions given in the discovery work are dynamically unstable on timescales of just hundreds of years. It is worth noting that repeating the dynamical stability testing with the much lower host-star mass proposed in Mortier et al. (2013) – resulting in proportionately lower planetary masses – yields the same degree of instability, and results that are indistinguishable from those presented in figure 2.

We have re-fit the observational data, and verified that the best fit system parameters are consistent with those reported by Niedzielski et al. (2009), and that the dynamically stable configurations are highly disfavoured by the observations. In light of our results, there is a clear need for more observational data to be obtained for this object over the coming years. As more data becomes available, covering a longer observational arc, the orbital parameters for the candidate planets will be refined, and it may be the case that alternative planetary solutions are found that represent a better fit to the data. For example, it may be that future observations reveal that the system hosts additional planets, which would significantly modify the architecture of the system from that proposed in the discovery work (e.g. Wittenmyer et al., 2013b; Wittenmyer et al., 2014). Follow-up observations are always important when studying exoplanetary systems – but this is particularly true for proposed systems that display large uncertainties in the best fit solution or exhibit significant dynamical instability.

Taken in concert, our results highlight how dynamical studies of exoplanetary systems in which multiple massive companions are proposed can act to provide significant additional constraints to the precision with which the orbits of the candidate planets can be determined. We have shown how such studies may be used to help differentiate between solutions of similar quality that provide significantly different orbital architectures – a result that is not that uncommon given that the chi-squared surfaces near a best-fit solution can often be very flat, and feature numerous degenerate secondary minima. It may often be the case that the orbital architecture resulting in the lowest chi-squared for a given planetary system is unphysical, when the dynamics of the system are taken into account, whilst the true solution lurks in a local chi-squared minimum that is not quite as deep – and dynamical integrations present the best available tool to resolve this dichotomy.

## Acknowledgements

The authors wish to thank the referee of this paper for providing swift and helpful feedback. The work was supported by iVEC through the use of advanced computing resources located at the Murdoch University, in Western Australia. This research has made use of NASA's Astrophysics Data System (ADS), TCH gratefully acknowledges financial support from the Korea Research Council for Fundamental Science and Technology (KRCF) through the Young Research Scientist Fellowship Program, and also the support of the Korea Astronomy and Space Science Institute (KASI) grant 2013-9-400-00. JPM is supported by Spanish grant AYA 2011/26202.